# Structural Study of a Self-Assembled Gold Mesocrystal Grain by Coherent X-ray Diffraction Imaging

Jerome Carnis,[†] Felizitas Kirner,[⊥] Dmitry Lapkin,[†] Sebastian Sturm,[¶] Young Yong Kim,[†] Igor A. Baburin,[§] Ruslan Khubbutdinov,[†,‡] Alexandr Ignatenko,[†] Ekaterina Iashina,[§] Alexander Mistonov,[§] Tristan Steegemans,[⊥] Thomas Wieck,[¶] Thomas Gemming,[¶] Axel Lubk,[¶] Sergey Lazarev,[†,&] Michael Sprung,[†] Ivan A. Vartanyants,[†,‡*] and Elena V. Sturm[⊥*]

Tuesday, March 2, 2021

[†] Deutsches Elektronen-Synchrotron DESY, Notkestrasse 85, D-22607 Hamburg, Germany

[⊥] University of Konstanz, Universitätsstraße 10, 78457 Konstanz, Germany

[¶] Leibniz Institute for Solid State and Materials Research, Helmholtzstraße 20, 01069 Dresden, Germany

[§] TU Dresden, Bergstraße 66b, 01062 Dresden, Germany

[‡] National Research Nuclear University MEPhI (Moscow Engineering Physics Institute), Kashirskoe shosse 31, 115409 Moscow, Russia

[§] Saint-Petersburg State University, University Embankment 7/9, 199034 St Petersburg, Russia

[&] Present address: Bruker AXS Advanced X-ray Solutions GmbH, Östliche Rheinbrückenstraße 49, 76187 Karlsruhe, Germany

# ABSTRACT

Mesocrystals are nanostructured materials consisting of individual nanocrystals having a preferred crystallographic orientation. On mesoscopic length scales, the properties of mesocrystals are strongly affected by structural heterogeneity. Here, we report the detailed structural characterization of a faceted mesocrystal grain self-assembled from 60 nm sized gold nanocubes. Using coherent X-ray diffraction imaging, we determined the structure of the mesocrystal with the resolution sufficient to resolve each gold nanoparticle. The reconstructed electron density of the gold mesocrystal reveals its intrinsic structural heterogeneity, including local deviations of lattice parameters, and the presence of internal defects. The strain distribution shows that the average superlattice obtained by angular X-ray cross-correlation analysis and the real, "multidomain" structure of a mesocrystal are very close to each other, with a deviation less than 10%. These results will provide an important impact to understanding of the fundamental principles of structuring and self-assembly including ensuing properties of mesocrystals.

Assembling of individual nanocrystals (NCs) in an organized superstructure offers the possibility of combining the physical properties of the individual NCs, like surface plasmon resonances or superparamagnetism, with other physical properties provided by the superstructure, such as mechanical stiffness or geometric coordination.[1,2] NCs can be self-assembled into a superstructure, which is also known as a mesocrystal, where the separate NCs share a common crystallographic orientation.[3-5] Among suitable building blocks for mesocrystals, gold nanoparticles have attracted special attention due to the prospect of utilizing their surface plasmon resonance properties. For example, a cancer-selective amplification of chemoradiation with plasmonic nanobubbles has been reported for clusters of gold nanoparticles.[6,7] Similarly, gold mesocrystals show high potential in the detection of traces of chemical species using surface-enhanced Raman scattering (SERS).[8-13]

In these applications, both the shape and size of the NCs as well as their mutual arrangement are highly influential on the plasmonic properties.[14] In particular, the crystallographic structure of the assemblies (mesocrystals) determines the coupling strength and coherent superposition of the NCs plasmon polariton modes, which leads to the measured enhancement of the electromagnetic field[13,15] and the emergence of hybridized plasmon bands.[16] Moreover, when anisotropic NCs are arranged in a mesocrystalline structure, coupling and amplification of directional physical properties can be expected.[4]

While the plasmonic properties of individual nanoparticles or two-dimensional (2D) assemblies can be easily characterized and related to the particle shape and size, the evaluation of optical responses from three-dimensional (3D) superstructures is not straight forward and requires exact knowledge not only of the internal symmetry, translational, and orientational order of the building blocks within the superlattice, but also structural heterogeneity (e.g. presence of defects). This knowledge is also important for the fundamental understanding of the phase behavior of nanocubes during the self-assembly process as well as the stability of superstructures.[17-19]

Previously, we presented a detailed structural characterization of PbS and iron oxide mesocrystalline superstructures using the combination of different transmission electron microscopy (TEM) techniques (including imaging, electron diffraction, high resolution TEM, electron tomography and electron holography).[20-25] TEM is capable to resolve very small structures, but quickly reaches an application limit for multilayered superstructures and larger mesocrystals, in particular if composed of more than 20 nm sized gold particles due to their strong electron scattering and absorption. In the few micrometers thick 2D layers of mesocrystals, the structural relationship between the atomic lattice and superlattice of nanocrystals can be studied by X-ray nanodiffraction and angular X-ray cross-correlation analysis (AXCCA) techniques.[26-30] However, to fully understand the mesocrystal formation process and to get insight into the fundamental principles of structure - property relationship of such complex material, detailed structural characterization in 3D is crucial. To study the structure of these materials in a non-destructive way and resolve potential defects, we use here coherent X-ray diffraction imaging (CXDI).[31,32]

The CXDI is a lens-less imaging technique that makes use of the coherence properties of the X-ray beam. When a finite object is illuminated by a coherent X-ray beam, interferences between the incoming wavefront and the scattered beams generate a diffraction pattern, which can be recorded in the far-field by a 2D detector with proper sampling.[33] By stepwise rotating the sample and recording 2D diffraction patterns, one can measure a full 3D diffraction pattern and then use iterative algorithms to determine the 3D electron density of a sample.[34,35] A complex amplitude object is reconstructed, whose modulus is directly related to the object's electron density in the forward scattering geometry.[33] This technique has already been applied successfully to a colloidal grain of silica nanospheres (diameter 230 nm), where the accuracy of the determined positions of each colloidal sphere in the lattice was ~9 nm.[36] Here, we push forward this approach to solve the structure of a gold mesocrystal grain self-assembled from 60 nm sized gold nanocubes.

**Results and Discussion**

The gold NCs stabilized by cetyltrimethylammonium chloride (CTAC) were synthesized in a three-step seed-mediated method,[21] purified by centrifugation and assembled to faceted mesocrystals (slightly distorted tetragonal prisms, see Figure 1) using depletion forces (see Methods for details). To perform a detailed structural characterization of the superstructure by CXDI, a rectangular cuboid with dimensions about 1.25 μm × 1.25 μm × 1.5 μm was cut from the central part of one of the grown mesocrystals using a focused ion beam (FIB) and was mounted on a tungsten micromanipulator tip (see Figure 2).

The CXDI measurements were performed at the Coherence Applications Beamline P10, at the PETRA III (DESY, Germany) storage ring. A sketch of the experimental setup is shown in Figure 2 (see Methods for details). The stack of 2D diffraction patterns measured in our experiment was interpolated into an orthonormal frame, resulting in a 3D reciprocal space map of our grain. An isosurface of this 3D diffraction pattern is shown in Figure 3a. One can distinguish streaks with interference fringes at low scattering angles in Figure 3b, which is due to interference of coherent X-rays on the opposite facets of the mesocrystal. In addition, several orders of superlattice reflections at larger wave vector transfers values can be observed. Between the Bragg peaks, we measured a complex speckle pattern, which encodes the local information about the relative positions of scatterers in the mesocrystal. The direct beam was masked, corresponding to the white area in the center of diffraction pattern as shown in Figure 3b. In Figure 3c, an angular average of the diffraction intensities is shown. Although we are dealing with a single-domain mesocrystal, Bragg peaks are broadened due to the highly defective structure of the ensemble. Near $q_z \sim \pm 0.1$ nm$^{-1}$, in Figure 3b, Bragg peaks are split, which suggests the presence of defects in the superlattice.[37] The first evaluation of the angular averaged X-ray diffraction profile shown in Figure 3c suggests a simple cubic symmetry (space group $Pm\bar{3}m$) of the superlattice with a lattice parameter $a \sim 62$ nm.

We further analyzed the 3D reciprocal space data by means of the AXCCA technique,[26-30] which can provide more information about the superlattice structure. This method was specifically modified to perform analysis in 3D reciprocal space. We calculated the cross-correlation functions (CCFs) $C(q_1,q_2,\Delta)$ between the most prominent peaks in the reciprocal space radial profile of Figure 3c, at momentum transfer values $q_1 = 0.104$ nm$^{-1}$, $q_2 = 0.144$ nm$^{-1}$, $q_3 = 0.172$ nm$^{-1}$, and $q_4 = 0.208$ nm$^{-1}$. The most representative CCFs are shown in Figure 4a-c. The CCFs reflect angular correlations between the Bragg peaks from the mesocrystalline lattice. The observed peaks in the CCFs do not perfectly fit the initially expected simple cubic lattice model of the mesocrystalline structure. Simulating peak positions for a triclinic lattice (space group $P1$) shown in Figure 4d, we found the optimal angles between the lattice basis vectors. Combining these data with the intensity profiles along the main crystallographic directions ([100], [010] and [001]) we obtain the following unit cell parameters in real space: $a = b = 63.2 \pm 0.1$ nm, $c = 62.2 \pm 0.1$ nm, $\alpha = \beta = 75 \pm 1°$, $\gamma = 90 \pm 1°$.

The anisotropy of the form-factor of the cubic NCs provides an additional information about the angular orientation of the NCs inside the superlattice. The form-factor maxima are located along the normal to the facets of the NCs. Thus, one can study the angular position of the NCs form-factor maxima with respect to the Bragg peaks and further determine the NC orientation with respect to the superlattice crystallographic axes in real space. The angular position of corresponding maxima was obtained by correlating the simulated form-factor for a cubic NC of the size of 59 ± 1 nm with the experimental intensities at $q = 0.477$ nm$^{-1}$. This $q$ value is the most suitable one to calculate correlations, since it corresponds to one of the maxima of the form-factor, and does not contain any structure factor features. As a result, we obtained an averaged real space model of the entire unit cell including the oriented NCs that is shown in Figure 4d. The facets of the NCs are aligned neither with the (001) plane, which is parallel to the substrate surface, nor with the [001] axis, but are tilted by an angle $\delta \approx 7°$ with respect to this axis. We would like to note here that reciprocal space analysis provides information about the average structure over the whole mesocrystalline grain only. Local structural features are resolved only by employing phase retrieval to reconstruct the electron density of the whole grain.

The 3D diffraction pattern was inverted to a real space 3D electron density map using iterative phase retrieval reconstruction algorithms,[31,32] providing a full overview of the internal structure of the mesocrystal (see for details Methods). An isosurface of the reconstructed mesocrystal together with 2D slices parallel to (100), (010) and (001) planes of the superlattice is shown in Figure 5. The first striking observation is that CXDI can fully resolve individual nanocubes. The voxel size of the reconstruction is 9.4 nm × 9.4 nm × 9.4 nm, while the resolution estimated by the phase-retrieval transfer function[38] (PRTF) is ~21 nm, which is smaller than the 60 nm size of the individual NCs. The reconstructed structure is tilted in the vertical direction, in agreement with the unit cell parameters of the superlattice obtained by the AXCCA.

The 2D slices of the electron density of the mesocrystalline grain (see Figure 5b-d) show the presence of several type of defects including point defects, lattice bending, crack, and voids. One could argue that the structure was damaged during sample preparation by the FIB. Indeed, it has been shown that FIB milling can induce a dislocation network in the outer layers of gold NCs.[39] However, this effect is expected to be limited to the first few tens of nanometers at the surface of the object, which does not explain the high inhomogeneity of the mesocrystal structure. Consequently, the defects likely formed during the crystal growth or the post-growth deformation processes (including the crystal contraction during the drying processes and solvent evaporation). Defects of both origins are important since application of mesocrystals often requires the material in a dry state.

It is also evident from the reconstructed electron density (see Figure 5), that the packing order of nanoparticles within (001) planes increases substantially in the first three layers from the substrate. We would like to note, that these (001) planes are parallel to the substrate and basal plane of mesocrystals and are perpendicular to the growth direction. In general, the ordering within (001) planes is also much higher than within (100) and (010) planes (see Figure 5). These findings are consistent with the layer growth mechanism of mesocrystals, proceeding by particle-by-particle attachment to the facet grown parallel to the substrate. In addition, during the drying process (e.g. solvent evaporation) the nonhomogeneous contraction of the mesocrystal occurs mainly perpendicular to the basal plane and additional shear stress induces the formation of additional cracks and voids, which mainly propagate perpendicular to substrate (see Figure 5). The most prominent here is a large crack through the entire mesocrystal, visible on the right side of Figure 5b,c, following the (010) plane, where its zigzag nature suggests a post-growth deformation during the drying process as origin of the crack formation.

To perform a more detailed analysis of the crystal's intrinsic heterogeneity, the 3D reconstructed electron density map was analyzed with a blob detection algorithm to extract the position of the individual NCs within the superlattice. The assigned position of individual NCs was further used to determine the local variations of packing order with respect to the average superlattice determined by the AXCCA. To do so, we aligned the averaged superlattice positions with the particle positions detected by the algorithm at the middle of the mesocrystal grain. The obtained displacement map clearly visualizes the lattice distortion across the entire 3D volume of the mesocrystal (Figure 6a). The shift of nanoparticle positions from their averaged value within the superlattice is most significant close to the structural defects (see Figure 5, for comparison). For example, the crack is again clearly visible at the right side of mesocrystal grain. Furthermore, one can see that for the first three layers from the substrate, the whole lattice plane is still significantly displaced in comparison to the average lattice. The upper left part of the crystal appears to be sheared approximately half a unit cell in a diagonal direction.

In Figure 6b, examples of extracted local primitive unit cells together with corresponding Voronoi–Dirichlet polyhedron (VDP), also known as the Wigner-Seitz cell for a 3D periodic lattice, at different positions within the mesocrystal are shown. The parameters and geometry of VDP constructed around the "central particle" of each selected cell illustrate the changes of the local symmetry in the arrangement of nearest neighbors (e.g. coordination shell) and coordination number (CN) equal to the number of facets.[40,41] The magnitude of the dimensionless second moment of inertia of the VDP, normalized to its volume (so-called $G_3$-factor) can be used to estimate the degree of distortion of the coordination shell (e.g. for cubes $G_3$ is equal to $1/12 \approx 0.0833$, for cuboctahedra $19/[192(2)^{1/3}] \approx 0.0785$, and for spheres $(1/5)(3/4\pi)^{2/3} \approx 0.0770$).[40,42] In comparison to cubic VDP of the simple cubic cell (CN is 6; number of vertices is 8, see Figure 6c), the more complex VDP of the averaged unit cell determined by AXCCA has 12 facets (CN = 12) and 18 vertices with $G_3 = 0.0810$ (Figure 6c). Interestingly, although the VDPs constructed for the several selected primitive cells (Figure 6b) have 14 facets (CN = 14), the $G_3$ parameter varies only from 0.0812 to 0.0831 (just between the values of a simple cubic and triclinic cells determined by the AXCCA).

It is worth to note, that the averaged AXCCA lattice can be seen as a distorted primitive cubic arrangement of the particles where each particle has 6 nearest neighbours and additionally 6 next-nearest neighbours, as evidenced by the faces of the VDP. The distortion can be understood as follows: the square layers of simple cubic lattice with the parameters $a = b = 62.2$ nm, $\gamma = 90°$ are sheared against each other. The distortions can be attributed both to the smoothed polyhedral shape of the particles that drives the whole arrangement in the direction of a close packing, and to the defects of the structure. Thus, it is not surprising, that the $G_3$ values calculated from experimental data are very close to the $G_3$ of the VDP in the simple cubic cell, showing that although the real symmetry of the superlattice cell is lower, its topology deviates only slightly from the simple cubic arrangement. This deviation might be also a result of the tilting of our slightly truncated gold NCs within the superlattice as indicated by the AXCCA (Figure 4d). This allows to achieve a more efficient space-filling arrangement in accordance with the so-called "bump-to-hollow" packing principle known for molecular crystals[43] and also reported for other mesocrystalline self-assembled structures.[18,22] In addition, we found, that this average primitive cell can actually be identified to have a higher symmetry and corresponds to a primitive cell of a centred monoclinic lattice that has the following parameters in the conventional setting *I*2/*m:* $a = 62.2$ nm, $b = 89.38$ nm, $c = 88.25$ nm, $\beta = 109.52°$. For convenience, however, we prefer to use the original primitive setting (see Figure 4d) in the whole manuscript. We also calculated the pair distribution function (PDF) based on the detected positions of individual nanocrystals within the superlattice (see Figure 6d). The PDF confirmed that the short-range order of the nanoparticles is close to an arrangement with the average cell, while the long-range order (at higher distances *r*) is significantly disturbed due to the presence of defects and lattice deformations in the real structure of the mesocrystal. The first peak of the PDF curve was fitted by a Gaussian function (see Figure 6d inset), revealing that the average distance to the neighboring particles of the first coordination shell is around 62.7 nm and that the precision of the detected particle positions (given by the standard deviation) is better than 5.9 nm, since the spread includes also the particle displacements due to lattice deformations.

Although the displacement map already gives a good understanding of the large lattice deformation across the whole mesocrystal volume, we additionally extracted a complete superlattice strain map. As a tool to describe the deformation of the superlattice, we defined the superlattice strain tensor as the deviation of the actually observed structure of a mesocrystal compared to the defect free lattice model (determined by the AXCCA) in analogy to the linear elasticity theory for atomic lattices[44]. This linear model, which neglects higher order derivatives, is of course only valid for small deformations. Although formally analogous to the strain as defined in linear elasticity theory (Hooke's law), herein it is merely used as a tool to describe the deviation of the observed experimental structure from the average model. Here, we are not aiming to distinguish an elastic and plastic deformation, nor can we say anything about the stress of the system. In addition, the superlattice dilatation (that is a sum of diagonal elements of the strain tensor) can easily be retrieved, giving a direct visualization of local lattice contractions and expansions (see Figure 7). Figure 7a,b illustrates the highly mosaic structure of the mesocrystal. Even though there are many local fluctuations in the crystal structure, there certainly seems to be a tendency of positive dilatation (lattice expansion) in the vicinity of the huge crack.

The individual strain $\varepsilon_{ij}$ and rotation $\omega_{ij}$ components can be found in Figure 7b. The most obvious feature is the "bipolar" structure, for example, in $\varepsilon_{xy}$ and $\omega_{yz}$, which is consistent with the lattice bending, and can be observed in the slices in Figure 5. The calculated strain tensor and rotation components are quantitatively summarized in the histograms (Figure 7c). The measured strain in all components is in a reasonable range of ±10%. The sharpest distribution is shown by the $\varepsilon_{yy}$ tensor component, which is closest to the $a_3$ growth direction of the mesocrystal. In order to verify the fidelity of our calculated superlattice strain, we reapplied the strain to an ideal crystal lattice (i.e. averaged superlattice determined by the AXCCA) and were able to reproduce most structural features of the experimentally measured mesocrystal.

**CONCLUSION**

In conclusion, we have used a combination of the angular X-ray cross-correlation analysis and coherent X-ray diffractive imaging, to study the 3D structure of a gold mesocrystal. The achieved resolution of 21 nm allowed imaging of individual nanocubes in the mesocrystal grain. Importantly, the precision of the detected particle positions was better than 6 nm. The results reveal a strongly defective structure of the mesocrystal with overall monoclinic symmetry. This technique shows a strong potential in systems that cannot be studied by classical optical or electron microscopy methods and contributes to a better understanding of structural features of self-assembled mesocrystals and especially intrinsic structural heterogeneity (including deviation of crystal symmetry, variation of lattice parameter, distribution of defects and strains). Importantly, the real mesocrystal structure was analyzed with unprecedented detail, calculating a displacement map and the entire strain tensor of the whole specimen.

The unexplored deformation behavior and related changes in structure and functional properties of nanoparticles superlattices limit their promising implementation into devices.[45] Therefore, the precise determination of their crystal structure and especially intrinsic structural heterogeneity is crucial for understanding this complex behavior. The proposed methodological approach could be used not only to perform such non-destructive structural characterization of nanostructured materials, but also to build more adequate structural models which in turn could serve as a basis for the precise prediction of the physical properties of nanoparticles self-assemblies using the computational methods.

## METHODS

### Synthesis of Gold Nanocubes

The synthesis of Gold Nanocubes was performed by a three-step seed-mediated growth method based on a previously described procedure.[21] First, small gold seeds are synthesized which are grown to spheres in a second step. These spheres are then processed for the synthesis of cubic particles. An aqueous solution of $HAuCl_4$ (1.00 mL, 0.01 M) and an aqueous solution of hexadecylpyridinium chloride (CPC) (50.00 mL, 0.10 M) were mixed in a glass vial (100 mL) and tempered to 27°C. An aqueous solution of ascorbic acid (0.75 mL, 0.10 M) was added, followed by a rapid injection of 400 µL of washed spheres. The solution turned pinkish red and was kept at 27°C for 3 h. The Gold Nanocubes are collected by centrifugation at 9 000 rpm for 5 min and redispersed in 0.02 M CPC solution. The cubes were characterized by TEM and UV/Vis, the single crystalline nature of the particles was proven using electron diffraction.

**Preparation of Gold Mesocrystals.**

An aqueous solution having approximately $10^{11}$ particles/mL and a CPC concentration of 0.02 M was prepared. 400 µL of this solution were filled in a 1 mL shell vial with a silicon wafer (7 × 5 mm$^2$). The solution was carefully overlayered with 400 µL of 25 wt.% cetyltrimethylammonium chloride (CTAC) in H$_2$O. After 12 h, the solution was carefully removed, and the silicon wafer was washed with acetone to yield mesocrystals.

**X-ray experiment**

Monochromatic X-rays of 8.7 keV were focused down to ~2.4 (H) × 2.0 (V) µm$^2$ at the sample position completely covering the mesocrystal grain. An electron microscopy image of the sample mounted on the tip is shown in Figure 2. The polymer tip was fixed on a rotation stage around the vertical axis *z*. At each angular position, the 2D far-field diffraction pattern was recorded by the Eiger4M detector positioned 4.95 m downstream from the sample. The sample was rotated by steps of 0.5° over a range of 180° and, by that, the full 3D diffraction pattern was measured. At each angular position, a series of 10 frames of 0.2 s exposure each were measured, corresponding to 2 s accumulated exposure to the non-attenuated X-ray beam, giving in total 18 min of measurements per sample. The sample was cooled using a liquid nitrogen cryostat, in order to avoid radiation damage of the organic ligand stabilizing NCs which could induce the NCs coalescence and destroy the superlattice ordering.

**Iterative phasing and resolution estimate**

Phase retrieval was carried out on the interpolated diffracted intensity data using PyNX package,[46] imposing at each iteration that the calculated Fourier intensity of the current object agrees with the measured data. The metric used to estimate the goodness of the fit during phasing was the free log-likelihood,[47] available in PyNX. Defective pixels for experimental data and gaps in the detector were not used for imposing the reciprocal space constraint mentioned above and thus were evolving freely during phasing.

The initial support was estimated from the autocorrelation function of the 3D diffraction intensity that included only the first superlattice Bragg peaks. For larger reciprocal space this support was correspondingly rescaled. While the iterative phase retrieval this support was evolved by application of the "shrink-wrap algorithm".[48] A series of 3600 Relaxed Averaged Alternating Reflections[49] plus 200 Error-Reduction[34] (ER) steps, including "shrink-wrap" algorithm[48] every 20 iterations were used. The phasing process included implementation of the Lucy-Richardson deconvolution that takes into account partial coherence effects.[50] The resulting point spread function is shown in Figure 8a-c of Supporting Information. To ensure the best reconstruction possible, we kept only the best 10 reconstructions from 1000 with random phase start and performed the mode decomposition.[47] The weight of the most prominent mode which was considered as a final result was 69%.

The resolution of the reconstruction was estimated using the normalized Phase-Retrieval Transfer Function[3] (PRTF) at a cutoff value of 1/e (see Figure 8d). The PRTF is a measure of how well the retrieved Fourier amplitudes match the square root of the measured diffraction intensity. After calculating the ratio of the reconstructed and measured amplitudes, the obtained 3D PRTF was azimuthally averaged over shells of constant $q$ and normalized to obtain the result shown in Figure 8d. Measured voxels of zero intensity were excluded from the PRTF calculation.

AUTHOR INFORMATION

**Corresponding Authors**

*Emails: Ivan A. Vartanyants: ivan.vartaniants@desy.de; Elena V. Sturm: elena.sturm@uni-konstanz.de

**Author Contributions**

I.A.V., E.V.S. designed the study. The samples were synthesized by F.K. The FIB cut preparation were performed by T.W., S.S., T.G. Scanning electron microscopy measurements were realized by F.K., S.S. Data acquisition was performed by J.C., Y.Y.K., R.K., A.I., E.I., A.M., S.L., M.S., I.A.V. The CXDI data analysis was performed by J.C. The AXCCA was performed by D.L. The structural analysis of reconstructed electron density was performed by S.S., I.A.B., A.L., E.V.S. The manuscript was written by J.C., F.K., D.L., S.S., A.L., I.A.B., I.A.V., E.V.S. All authors have given approval to the final version of the manuscript.

J.C. & F.K.; D.L. & S.S. contributed equally to this work.

ACKNOWLEDGMENTS

We acknowledge DESY (Hamburg, Germany), a member of the Helmholtz Association HGF, for the provision of experimental facilities. Parts of this research were carried out at PETRA III synchrotron facility and we would like to thank the beamline staff for assistance in using Coherence Application P10 beamline. This work was supported by the Helmholtz Associations Initiative Networking Fund (grant No. HRSF-0002) and the Russian Science Foundation (grant No. 18-41-06001); DFG (Deutsche Forschungsgemeinschaft) within the SFB 1214 (TP: B1). J.C. would like to thank Vincent Favre-Nicolin for his continuous improvement of the PyNX package used here for phase retrieval. S.S. and A.L. acknowledge funding from the European Research Council (ERC) under the Horizon 2020 research and innovation program of the European Union (grant agreement number 715620). S.L. acknowledges Competitiveness Enhancement Program

Grant of Tomsk Polytechnic University. E.S. thanks Res. Prof. Alexander Shtukenberg for the fruitful discussion and the Zukunftskolleg at the University of Konstanz and "Konstanzia Transition" program of equal opportunity council for financial support. The authors acknowledge support of the project by Prof. Edgar Weckert and careful reading of the manuscript by Dr. Thomas Keller.

**REFERENCES**

1. Zhou, L.; O'Brien, P., Mesocrystals - Properties and Applications. *J. Phys. Chem. Lett.* **2012,** *3*, 620-8.
2. Talapin, D. V.; Lee, J. S.; Kovalenko, M. V.; Shevchenko, E. V., Prospects of Colloidal Nanocrystals for Electronic and Optoelectronic Applications. *Chem. Rev.* **2010,** *110*, 389-458.
3. Cölfen, H.; Mann, S., Higher-Order Organization by Mesoscale Self-Assembly And Transformation of Hybrid Nanostructures. *Angew. Chem. Int. Ed.* **2003,** *42*, 2350-65.
4. Sturm, E. V.; Cölfen, H., Mesocrystals: Structural and Morphogenetic Aspects. *Chem. Soc. Rev.* **2016,** *45*, 5821-5833.
5. Sturm, E. V.; Cölfen, H., Mesocrystals: Past, Presence, Future. *Crystals* **2017,** *7*, 207.
6. Lukianova-Hleb, E. Y.; Ren, X.; Sawant, R. R.; Wu, X.; Torchilin, V. P.; Lapotko, D. O., On-Demand Intracellular Amplification of Chemoradiation with Cancer-Specific Plasmonic Nanobubbles. *Nat. Med.* **2014,** *20*, 778-784.
7. Lukianova-Hleb, E. Y.; Ren, X.; Zasadzinski, J. A.; Wu, X.; Lapotko, D. O., Plasmonic Nanobubbles Enhance Efficacy and Selectivity of Chemotherapy Against Drug-Resistant Cancer Cells. *Adv. Mater.* **2012,** *24*, 3831-7.
8. Kim, B.; Tripp, S. L.; Wei, A., Self-Organization of Large Gold Nanoparticle Arrays. *J. Am. Chem. Soc.* **2001,** *123*, 7955-6.
9. Wei, A.; Kim, B.; Sadtler, B.; Tripp, S. L., Tunable Surface-Enhanced Raman Scattering from Large Gold Nanoparticle Arrays. *Chemphyschem* **2001,** *2*, 743-5.
10. Zhu, Z.; Meng, H.; Liu, W.; Liu, X.; Gong, J.; Qiu, X.; Jiang, L.; Wang, D.; Tang, Z., Superstructures and SERS Properties o Gold Nanocrystals with Different Shapes. *Angew. Chem. Int. Ed.* **2011,** *50*, 1593-6.
11. Alvarez-Puebla, R. A.; Agarwal, A.; Manna, P.; Khanal, B. P.; Aldeanueva-Potel, P.; Carbo-Argibay, E.; Pazos-Perez, N.; Vigderman, L.; Zubarev, E. R.; Kotov, N. A.; Liz-Marzan, L. M., Gold Nanorods 3D-Supercrystals As Surface Enhanced Raman Scattering Spectroscopy Substrates for the Rapid Detection of Scrambled Prions. *PNAS* **2011,** *108*, 8157-61.
12. Matricardi, C.; Hanske, C.; Garcia-Pomar, J. L.; Langer, J.; Mihi, A.; Liz-Marzan, L. M., Gold Nanoparticle Plasmonic Superlattices as Surface-Enhanced Raman Spectroscopy Substrates. *ACS Nano* **2018,** *12*, 8531-8539.
13. Bian, K.; Schunk, H.; Ye, D.; Hwang, A.; Luk, T. S.; Li, R.; Wang, Z.; Fan, H., Formation of Self-Assembled Gold Nanoparticle Supercrystals with Facet-Dependent Surface Plasmonic Coupling. *Nat. Commun.* **2018,** *9*, 2365.

14. Ma, Y.; Li, W.; Cho, E. C.; Li, Z.; Yu, T.; Zeng, J.; Xie, Z.; Xia, Y., Au@Ag Core-Shell Nanocubes with Finely Tuned and Well-Controlled Sizes, Shell Thicknesses, and Optical Properties. *ACS Nano* **2010,** *4*, 6725-34.
15. Hamon, C.; Sanz-Ortiz, M. N.; Modin, E.; Hill, E. H.; Scarabelli, L.; Chuvilin, A.; Liz-Marzan, L. M., Hierarchical Organization and Molecular Diffusion in Gold Nanorod/Silica Supercrystal Nanocomposites. *Nanoscale* **2016,** *8*, 7914-22.
16. Sun, L.; Lin, H.; Kohlstedt, K. L.; Schatz, G. C.; Mirkin, C. A., Design Principles for Photonic Crystals Based on Plasmonic Nanoparticle Superlattices. *PNAS* **2018,** *115*, 7242-7247.
17. Smallenburg, F.; Filion, L.; Marechal, M.; Dijkstra, M., Vacancy-Stabilized Crystalline Order in Hard Cubes. *PNAS* **2012,** *109*, 17886-90.
18. Rossi, L.; Soni, V.; Ashton, D. J.; Pine, D. J.; Philipse, A. P.; Chaikin, P. M.; Dijkstra, M.; Sacanna, S.; Irvine, W. T., Shape-Sensitive Crystallization in Colloidal Superball Fluids. *PNAS* **2015,** *112*, 5286-90.
19. Kapuscinski, M.; Agthe, M.; Lv, Z. P.; Liu, Y.; Segad, M.; Bergström, L., Temporal Evolution of Superlattice Contraction and Defect-Induced Strain Anisotropy in Mesocrystals during Nanocube Self-Assembly. *ACS Nano* **2020,** *14*, 5337-5347.
20. Simon, P.; Rosseeva, E.; Baburin, I. A.; Liebscher, L.; Hickey, S. G.; Cardoso-Gil, R.; Eychmuller, A.; Kniep, R.; Carrillo-Cabrera, W., Pbs-Organic Mesocrystals: the Relationship Between Nanocrystal Orientation and Superlattice Array. *Angew. Chem. Int. Ed.* **2012,** *51*, 10776-81.
21. Kirner, F.; Potapov, P.; Schultz, J.; Geppert, J.; Müller, M.; González-Rubio, G.; Sturm, S.; Lubk, A.; Sturm, E., Additive-Controlled Synthesis of Monodisperse Single Crystalline Gold Nanoparticles: Interplay of Shape and Surface Plasmon Resonance. *J. Mater. Chem. C* **2020,** *8*, 10844-10851.
22. Brunner, J.; Baburin, I. A.; Sturm, S.; Kvashnina, K.; Rossberg, A.; Pietsch, T.; Andreev, S.; Sturm née Rosseeva, E.; Cölfen, H., Self-Assembled Magnetite Mesocrystalline Films: Toward Structural Evolution from 2D to 3D Superlattices. *Adv. Mater. Interfaces* **2017,** *4*, 1600431.
23. Brunner, J. J.; Krumova, M.; Cölfen, H.; Sturm, E. V., Magnetic Field-Assisted Assembly of Iron Oxide Mesocrystals: a Matter of Nanoparticle Shape and Magnetic Anisotropy. *Beilstein J. Nanotechnol.* **2019,** *10*, 894-900.
24. Sturm, S.; Siglreitmeier, M.; Wolf, D.; Vogel, K.; Gratz, M.; Faivre, D.; Lubk, A.; Buchner, B.; Sturm, E. V.; Cölfen, H., Magnetic Nanoparticle Chains in Gelatin Ferrogels: Bioinspiration from Magnetotactic Bacteria. *Adv. Funct. Mater.* **2019,** *29*.
25. Simon, P.; Bahrig, L.; Baburin, I. A.; Formanek, P.; Roder, F.; Sickmann, J.; Hickey, S. G.; Eychmuller, A.; Lichte, H.; Kniep, R.; Rosseeva, E., Interconnection of Nanoparticles within 2D Superlattices of Pbs/Oleic Acid Thin Films. *Adv. Mater.* **2014,** *26*, 3042-9.
26. Zaluzhnyy, I. A.; Kurta, R. P.; André, A.; Gorobtsov, O. Y.; Rose, M.; Skopintsev, P.; Besedin, I.; Zozulya, A. V.; Sprung, M.; Schreiber, F., Quantifying Angular Correlations between the Atomic Lattice and the Superlattice of Nanocrystals Assembled with Directional Linking. *Nano Lett.* **2017,** *17*, 3511-3517.
27. Mukharamova, N.; Lapkin, D.; Zaluzhnyy, I. A.; Andre, A.; Lazarev, S.; Kim, Y. Y.; Sprung, M.; Kurta, R. P.; Schreiber, F.; Vartanyants, I. A.; Scheele, M., Revealing Grain Boundaries and Defect Formation in Nanocrystal Superlattices by Nanodiffraction. *Small* **2019,** *15*, e1904954.
28. Maier, A.; Lapkin, D.; Mukharamova, N.; Frech, P.; Assalauova, D.; Ignatenko, A.; Khubbutdinov, R.; Lazarev, S.; Sprung, M.; Laible, F.; Löffler, R.; Previdi, N.; Bräuer, A.; Günkel,

T.; Fleischer, M.; Schreiber, F.; Vartanyants, I. A.; Scheele, M., Structure-Transport Correlation Reveals Anisotropic Charge Transport in Coupled PbS Nanocrystal Superlattices. *Adv. Mater.* **2020,** *32*, e2002254.
29. Kurta, R. P.; Altarelli, M.; Vartanyants, I. A., Structural Analysis by X-Ray Intensity Angular Cross Correlations. *Adv. Chem. Phys.* **2016,** *161*, 1-39.
30. Zaluzhnyy, I. A.; Kurta, R. P.; Scheele, M.; Schreiber, F.; Ostrovskii, B. I.; Vartanyants, I. A., Angular X-Ray Cross-Correlation Analysis (AXCCA): Basic Concepts and Recent Applications to Soft Matter and Nanomaterials. *Materials* **2019,** *12*, 3464.
31. Miao, J. W.; Charalambous, P.; Kirz, J.; Sayre, D., Extending the Methodology of X-Ray Crystallography to Allow Imaging of Micrometre-Sized Non-Crystalline Specimens. *Nature* **1999,** *400*, 342-344.
32. Pfeifer, M. A.; Williams, G. J.; Vartanyants, I. A.; Harder, R.; Robinson, I. K., Three-Dimensional Mapping of a Deformation Field Inside a Nanocrystal. *Nature* **2006,** *442*, 63-6.
33. Vartanyants, I. A.; Yefanov, O. M., Coherent X-ray Diffraction Imaging of Nanostructures. In *X-ray Diffraction. Modern Experimental Techniques.*, Seeck, O. H.; Murphy, B. M., Eds. Pan Stanford Publishing: Singapore 2015; pp 341-384.
34. Fienup, J. R., Phase Retrieval Algorithms: a Comparison. *Appl. Opt.* **1982,** *21*, 2758-2769.
35. Marchesini, S., Invited Article: A Unified Evaluation of Iterative Projection Algorithms for Phase Retrieval. *Rev. Sci. Instrum.* **2007,** *78*, 011301.
36. Shabalin, A. G.; Meijer, J. M.; Dronyak, R.; Yefanov, O. M.; Singer, A.; Kurta, R. P.; Lorenz, U.; Gorobtsov, O. Y.; Dzhigaev, D.; Kalbfleisch, S.; Gulden, J.; Zozulya, A. V.; Sprung, M.; Petukhov, A. V.; Vartanyants, I. A., Revealing Three-Dimensional Structure of an Individual Colloidal Crystal Grain by Coherent X-Ray Diffractive Imaging. *Phys. Rev. Lett.* **2016,** *117*, 138002.
37. Dupraz, M.; Beutier, G.; Rodney, D.; Mordehai, D.; Verdier, M., Signature of Dislocations and Stacking Faults of Face-Centred Cubic Nanocrystals in Coherent X-Ray Diffraction Patterns: a Numerical Study. *J. Appl. Crystallogr.* **2015,** *48*, 621-644.
38. Chapman, H. N.; Barty, A.; Marchesini, S.; Noy, A.; Hau-Riege, S. P.; Cui, C.; Howells, M. R.; Rosen, R.; He, H.; Spence, J. C.; Weierstall, U.; Beetz, T.; Jacobsen, C.; Shapiro, D., High-Resolution *ab initio* Three-Dimensional X-Ray Diffraction Microscopy. *J. Opt. Soc. Am. A Opt. Image Sci. Vis.* **2006,** *23*, 1179-200.
39. Hofmann, F.; Tarleton, E.; Harder, R. J.; Phillips, N. W.; Ma, P. W.; Clark, J. N.; Robinson, I. K.; Abbey, B.; Liu, W.; Beck, C. E., 3d Lattice Distortions and Defect Structures in Ion-Implanted Nano-Crystals. *Sci. Rep.* **2017,** *7*, 45993.
40. Blatov, V. A., Voronoi–Dirichlet Polyhedra in Crystal Chemistry: Theory and Applications. *Crystallogr. Rev.* **2004,** *10*, 249-318.
41. Blatov, V. A.; Shevchenko, A. P.; Proserpio, D. M., Applied Topological Analysis of Crystal Structures with the Program Package ToposPro. *Cryst. Growth Des.* **2014,** *14*, 3576-3586.
42. Peresypkina, E. V.; Blatov, V. A., Methods for Assessing the Degree of Sphericity of Molecules and a Study of Molecular Shapes in the Structure of Binary Inorganic Compounds. *Russ. J. Inorg. Chem.* **2003,** *48*, 237-245.
43. Kitaigorodkij, A. I., *Organic Chemical Crystallography.* 1961; Vol. Consultants Bureau New York p541.
44. Landau, L. D.; Pitaevskii, L. P.; Kosevich, A. M.; Lifshitz, E. M., *Theory of Elasticity*. 3rd ed.; Elsevier Science: 2012; Vol. 7.

45.	Giuntini, D.; Zhao, S.; Krekeler, T.; Li, M.; Blankenburg, M.; Bor, B.; Schaan, G.; Domènech, B.; Müller, M.; Scheider, I.; Ritter, M.; Schneider, G. A., Defects and Plasticity in Ultrastrong Supercrystalline Nanocomposites. *Science Advances* **2021,** *7*, eabb6063.
46.	Mandula, O.; Elzo Aizarna, M.; Eymery, J.; Burghammer, M.; Favre-Nicolin, V., PyNX. Ptycho: a Computing Library for X-Ray Coherent Diffraction Imaging of Nanostructures. *J. Appl. Crystallogr.* **2016,** *49*, 1842-1848.
47.	Favre-Nicolin, V.; Leake, S.; Chushkin, Y., Free Log-Likelihood As an Unbiased Metric for Coherent Diffraction Imaging. *Sci. Rep.* **2020,** *10*, 1-8.
48.	Marchesini, S.; He, H.; Chapman, H. N.; Hau-Riege, S. P.; Noy, A.; Howells, M. R.; Weierstall, U.; Spence, J. C. H., X-Ray Image Reconstruction from a Diffraction Pattern Alone. *Phys. Rev. B: Condens. Matter* **2003,** *68*, 140101.
49.	Luke, D. R., Relaxed Averaged Alternating Reflections for Diffraction Imaging. *Inverse Probl.* **2004,** *21*, 37.
50.	Clark, J. N.; Huang, X.; Harder, R.; Robinson, I. K., High-Resolution Three-Dimensional Partially Coherent Diffraction Imaging. *Nat. Commun.* **2012,** *3*, 993.

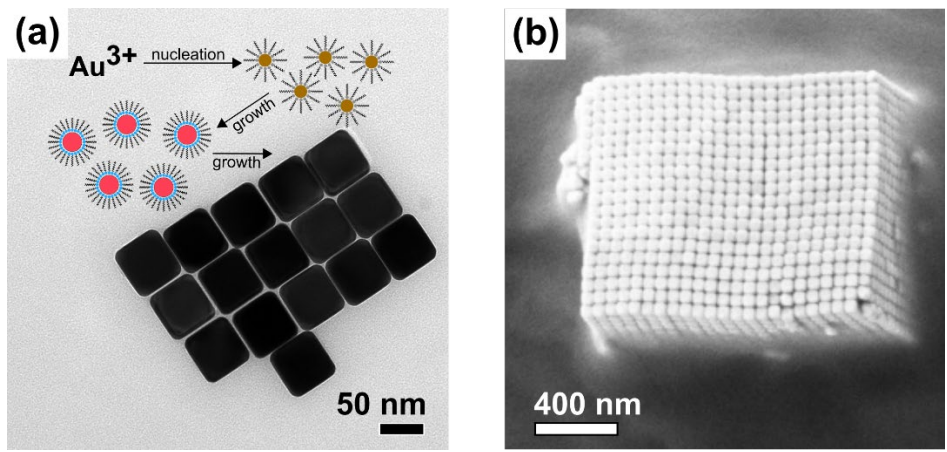

**Figure 1**. Synthesis and the preparation of gold mesocrystals. (a) TEM image of gold nanocubes stabilized by CTAC and synthesized using a seed-mediated approach. Electron diffraction pattern (bottom left) shows the perfect orientational alignment of the nanocubes along the [100] crystallographic direction within the assembly. (b) SEM image of the self-assembled gold mesocrystal prior to FIB preparation for the CXDI measurements.

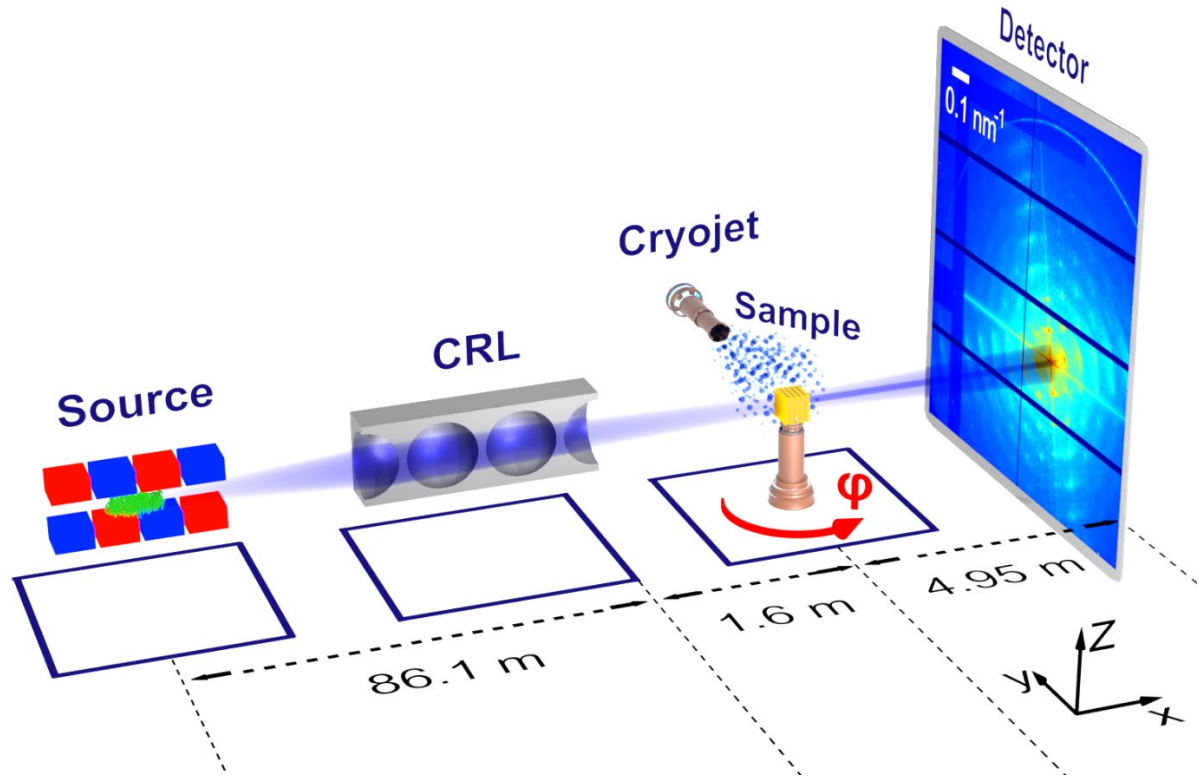

**Figure 2**. Schematic layout of the experiment. A monochromatic X-ray beam of 8.7 keV from the undulator source is focused by the Compound Refractive Lenses (CRL) to a spot larger than the mesocrystal grain. The sample was cooled using a liquid nitrogen cryostat. The far field diffraction patterns were measured by a photon counting Eiger4M detector positioned downstream the sample. The 3D diffraction map is obtained by rotating the sample around the vertical axis. In the inset, an SEM image of the mesocrystal grain is shown. The laboratory frame convention for the coordinate frame is shown in the figure.

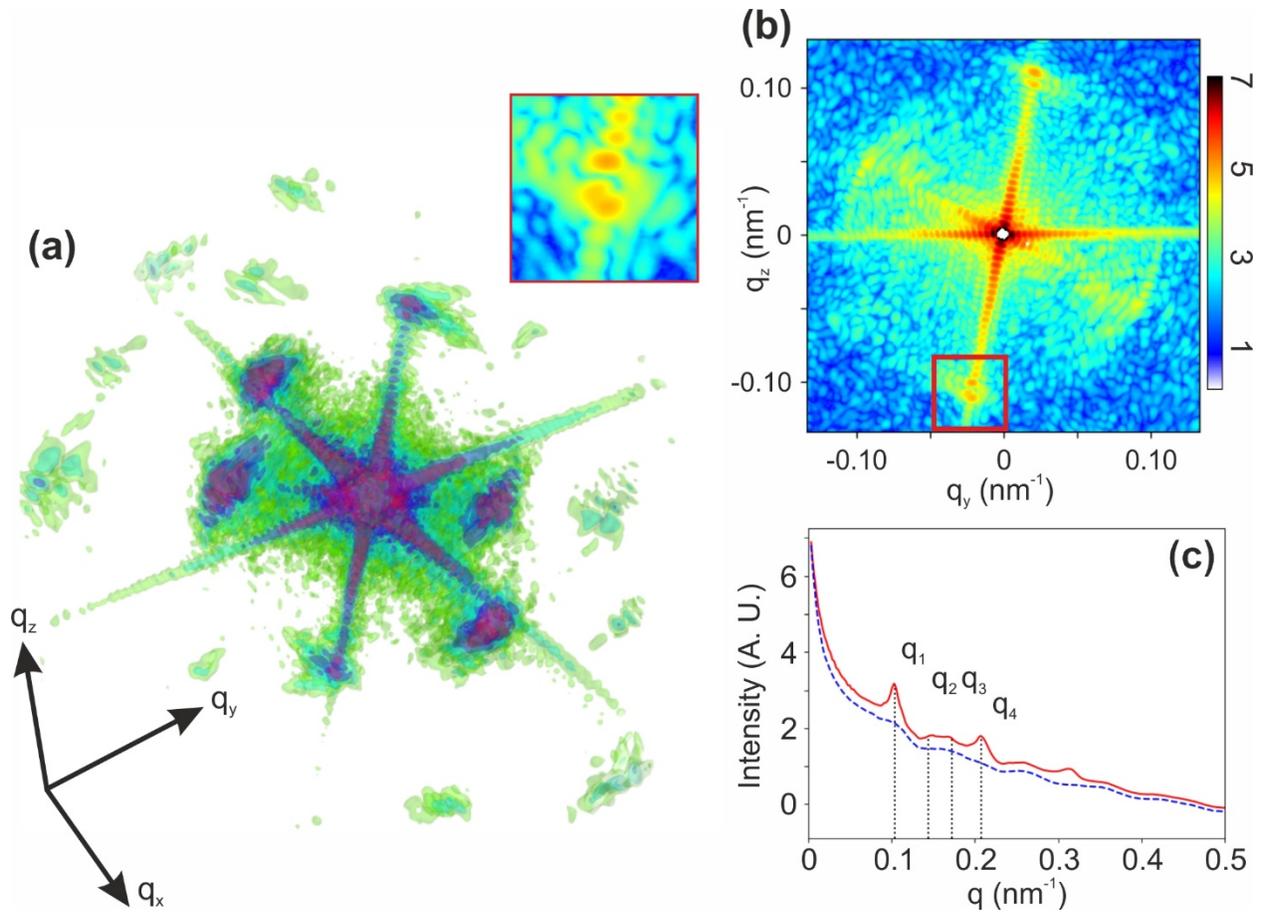

**Figure 3**. (a) Isosurface view (from 54% to 72% level) of the 3D diffraction pattern from the mesocrystalline grain shown in SEM image in the inset of Figure 2. The data has been gridded onto the orthonormal laboratory frame. The length of coordinate arrows corresponds to 0.1 nm$^{-1}$. (b) Slice at the center of the interpolated diffraction pattern in the $q_y q_z$ plane, showing the low-angle scattering region up to the first superlattice Bragg peaks. An enlarged view of the area outlined by a red box is shown in panel (a). (c) Intensity distribution as a function of momentum transfer value $q$ obtained by angular averaging of the 3D diffraction pattern (red line). The blue dashed line corresponds to the median value at momentum transfer q determined from the angular averaged values. The AXCCA was performed on the data obtained by subtracting these median values from the 3D diffraction pattern shown in (a). The vertical dotted lines correspond to the momentum transfer values used in the AXCCA: $q_1 = 0.104$ nm$^{-1}$, $q_2 = 0.144$ nm$^{-1}$, $q_3 = 0.172$ nm$^{-1}$, and $q_4 = 0.208$ nm$^{-1}$.

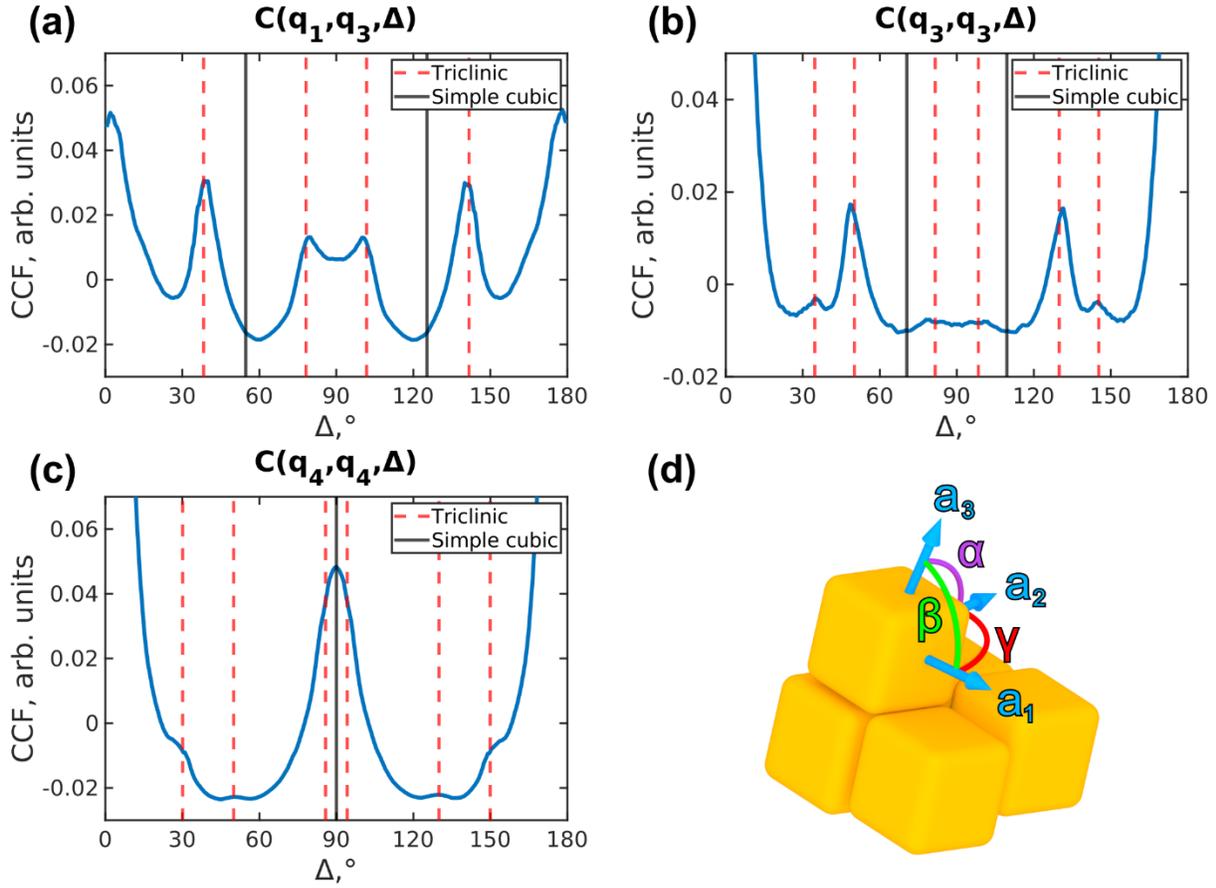

**Figure 4.** Cross-correlation functions (CCFs) C($q_1$, $q_2$, $\Delta$) calculated at the following momentum transfer values (a) $q_1 = 0.104$ nm$^{-1}$ and $q_3 = 0.172$ nm$^{-1}$, (b) $q_3 = 0.172$ nm$^{-1}$, and (c) $q_4 = 0.208$ nm$^{-1}$. The corresponding peak positions for the optimized triclinic unit cell are shown with red dashed lines. The peak positions for a simple cubic unit cell are shown with black lines. (d) A real space model of the average superlattice triclinic unit cell, with $a = b = 63.2 \pm 0.1$ nm, $c = 62.2 \pm 0.1$ nm, $\alpha = \beta = 75 \pm 1°$, $\gamma = 90 \pm 1°$. The orientation of the nanocubes within the superlattice is revealed by analysis of the anisotropic form-factor.

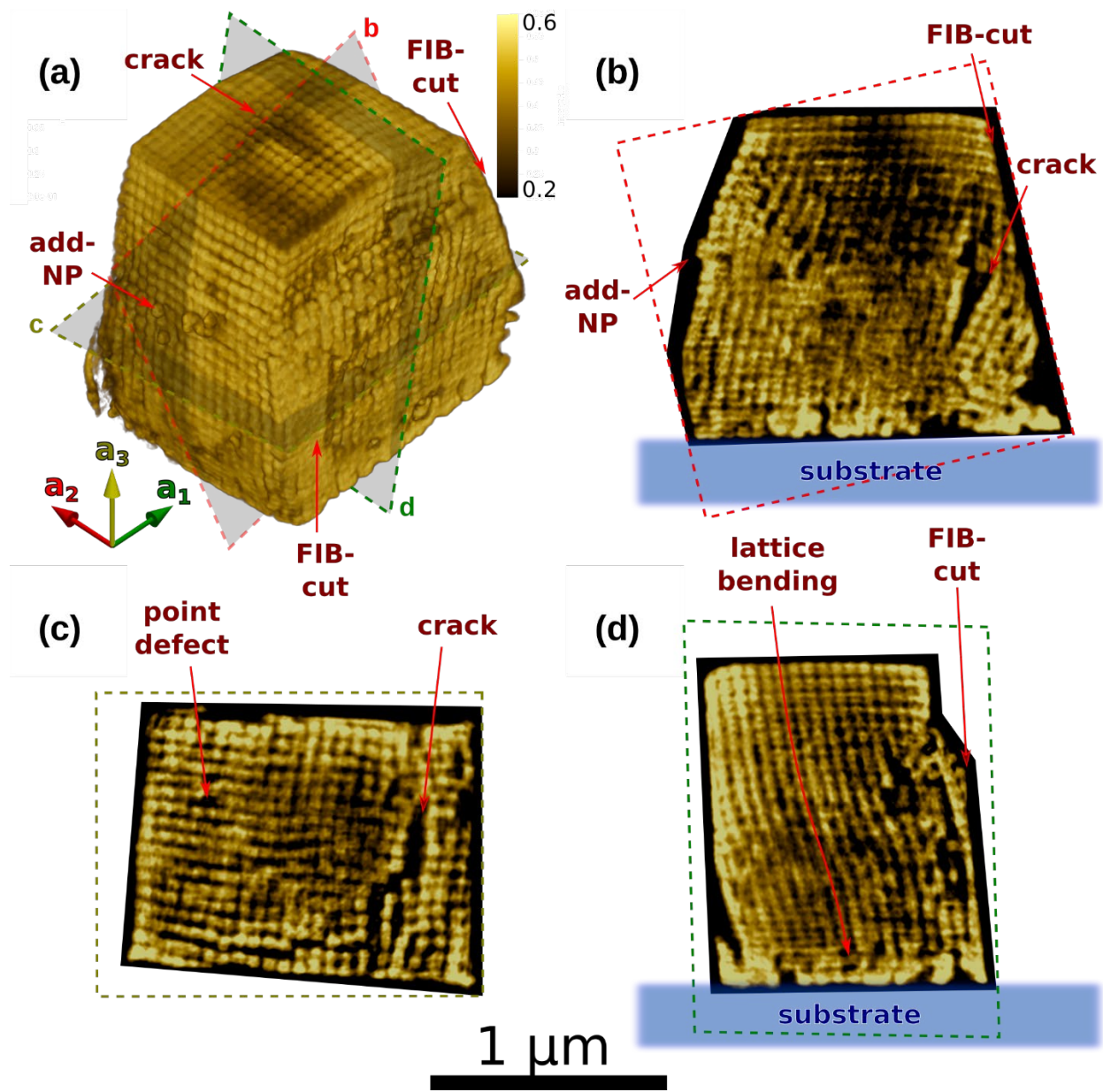

**Figure 5.** (a) Volume rendering of the reconstructed 3D electron density with the position of 2D slices used further to analyze the structure. (b-d) 2D slices parallel to (010) (b), (001) (c) and (100) (d) crystallographic planes through the center of the reconstructed mesocrystal with highlighted superlattice defects.

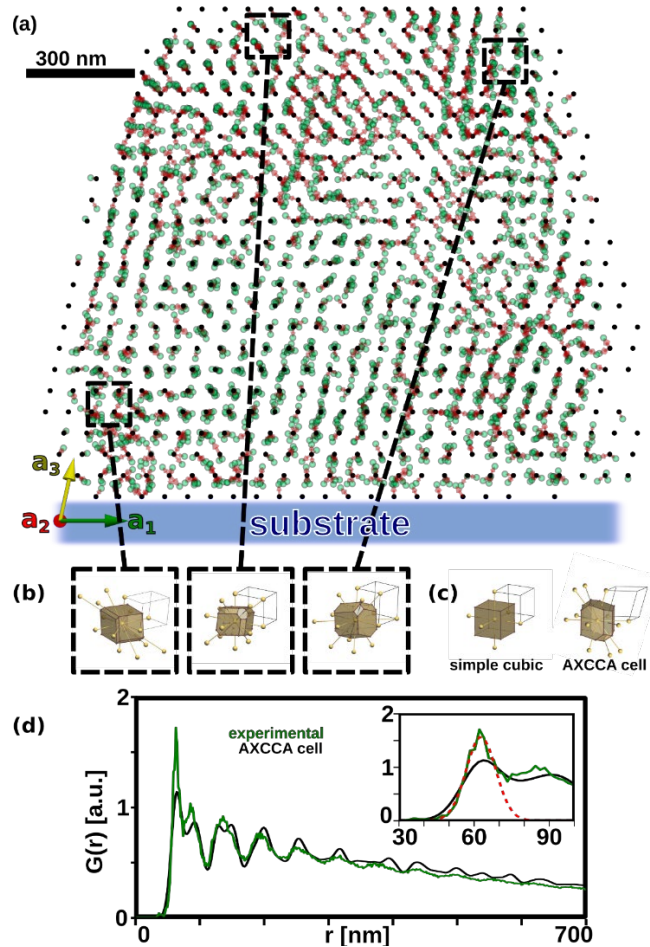

**Figure 6.** (a) Displacement map (red arrows) of the 3D superlattice, obtained based on positions of detected nanoparticles (green dots) with respect to an average superlattice (black dots) with unit cell parameters determined by the AXCCA. The two lattices were aligned at the center of the grain (b) Selected Voronoi–Dirichlet polyhedrons illustrating the change of local symmetry of the arrangement and the number of coordinating particles ($G_3$ from left to right: 0.0831, 0.0820, 0.0813). (c) Constructed Voronoi–Dirichlet polyhedrons of simple cubic lattice ($G_3 = 0.0833$) and the average superlattice unit cell determined by the AXCCA ($G_3 = 0.0811$). (d) Pair-distribution functions (PDF) obtained for two datasets. First, based on the positions of the detected nanoparticles shown in panel (a) by green dots (green line) and, second, based on an average superlattice unit cell determined by the AXCCA shown by black dots in panel (a) (black line). In the inset an enlarged part of the first two peaks of the PDF-function is shown. The first peak of the experimental curve was fitted by a Gaussian function (red line) which has its maximum position at 62.7 nm and standard deviation 5.9 nm.

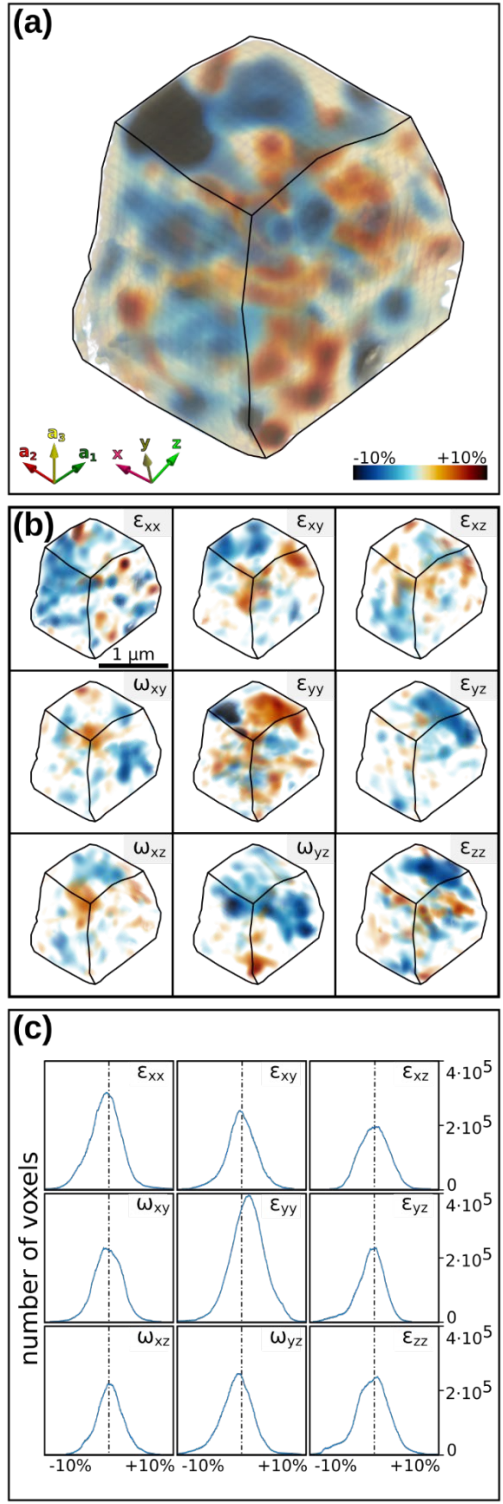

**Figure 7.** (a) Volume rendering of a 3D dilatation map of the superlattice. (b) Volume rendering of 3D maps of the strain tensor components ($\varepsilon_{xx} - \varepsilon_{zz}$) and rotations ($\omega_{xy} - \omega_{yz}$). The magnitude of strain is illustrated by the color scale. (c) Histograms of the strain tensor components and rotations.

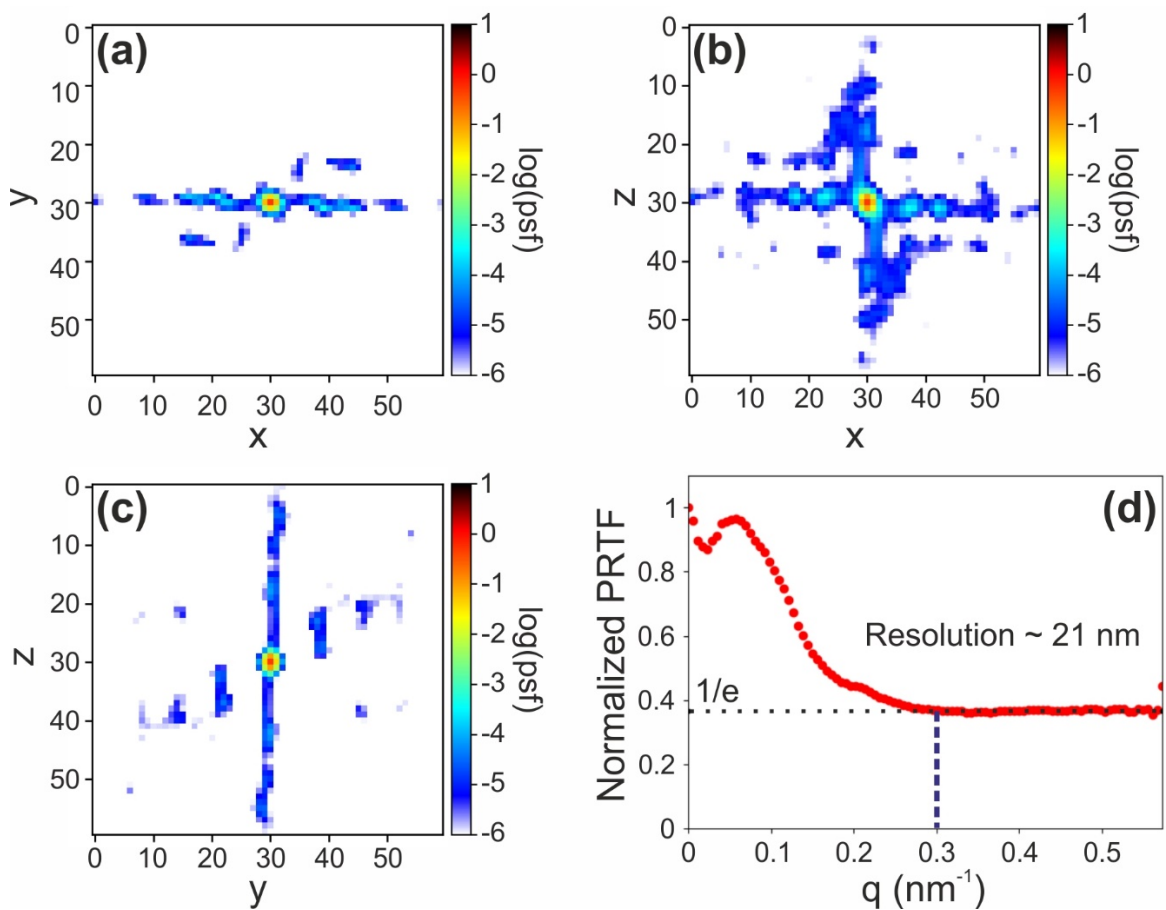

**Fig. 8**. (a-c) Slices through the point spread function obtained from application of the Lucy-Richardson algorithm, (d) Azimuthally averaged and normalized PRTF. The resolution is determined as the cross-over between the PRTF and the 1/e line.